  \documentclass[twocolumn,prl,aps]{revtex4}

 \usepackage[dvips]{graphicx}
 \usepackage[dvips]{graphics}

\newlength{\bxwidth}\bxwidth=0.8\textwidth

\begin{document}
\title{Phonon Mechanism of the Ferromagnetic Transition in ${La_{1-x}Sr_{x}MnO_{3}}$}
\author{D. Reznik$^{1,2\ast}$ and W. Reichardt$^1$}

\affiliation{$^1$Forschungszentrum Karlsruhe, Institut f\"ur Festk\"orperphysik, Postfach 3640, D-76021 Karlsruhe, Germany.\\
$^2$Laboratoire L\'eon Brillouin, CEA-CNRS, CE-Saclay, 91191 Gif sur Yvette, France.\\
}

% 25.40.Fq Inelastic neutron scattering
% 74.72.Bk Superconducting materials Y-based cuprates
% 74.25.Ha Superconductivity Magnetic properties

\begin{abstract}

Temperature dependence of longitudinal optical phonons with oxygen character was measured in ${La_{1-x}Sr_{x}MnO_{3}}$ (x=0.2, 0.3) by inelastic neutron scattering in the (1 0 0) cubic direction. The zone center mode intensity is consistent with the Debye-Waller factor. However the intensity of the bond-stretching phonons half way to the zone boundary and near the zone boundary decreases dramatically as the temperature increases through the ferromagnetic (FM) transition. We found evidence that the lost phonon spectral weight might shift into polaron scattering at the same wavevectors. The temperature evolution starts well below the onset of the FM transition suggesting that the transition is driven by phonon renormalization rather than by magnetic fluctuations.

\end{abstract}

\maketitle

Interplay between the charge, spin, and lattice degrees of freedom in perovskite manganites results in a multitude of unconventional properties of fundamental as well as practical interest. Ferromagnetic alignment of Mn core spins appears at some stoichiometries due to the double exchange interaction and the ferromagnetic-paramagnetic (FM) transition in these systems is accompanied by large magnetoresistance (MR). As predicted by theory \cite{Millis}, polarons forming and condensing above the (FM) transition temperature, T$_c$, have been observed by neutron scattering in ${La_{0.7}Ca_{0.3}MnO_{3}}$ and other manganites exhibiting Colossal Magnetoresistance (CMR) \cite{Adams}. Trapping of conduction electrons by these polarons above T$_c$ is held responsible for the CMR effect. In ${La_{0.7}Sr_{0.3}MnO_{3}}$, where the resistivity has a much smaller jump at T$_c$, the prevailing view is that double-exchange interaction alone is responsible for the essential physics \cite{Urushibara}. The current study of oxygen vibrations in ${La_{0.7}Sr_{0.3}MnO_{3}}$ and ${La_{0.8}Sr_{0.2}MnO_{3}}$ challenges this notion providing strong evidence for a lattice-driven mechanism of the FM transition in all ferromagnetic manganites. 

${La_{1-x}Sr_{x}MnO_{3}}$ (x=0.2, 0.3) have a cubic perovskite structure except for a rotation of the ${MnO_{6}}$ octahedra around the 111 direction, which makes it rhombohedral \cite{Dabrowski}. In addition, the crystals are twinned, so one always has to consider the effect of the superposition of the two domains. To simplify the analysis of our results we will use the cubic notation and treat the rotation of the octahedra as a rhombohedral distortion. The current study focused on longitudinal branches in the (1 0 0) direction, which is the same for the two domains, and there are no complications due to twinning.

The experiments were carried out on the triple-axis spectrometer 1T located at the ORPHEE reactor using doubly focusing Cu111 and Cu220 monochromator crystals and PG002 analyzer fixed at 13.7, 14.8, 30.5, or 35meV. Our samples were two high quality single crystals of ${La_{0.7}Sr_{0.3}MnO_{3}}$ and one single crystal of ${La_{0.8}Sr_{0.2}MnO_{3}}$ with the FM transition temperatures measured at 355K and 300K respectively.  The volume of each crystal was ~0.5cm$^3$. The results were the same for the two ${La_{0.7}Sr_{0.3}MnO_{3}}$ crystals. The neutron scattering spectra were divided by the Bose factor in order to compare one-phonon scattering intensities at different temperatures. 

We first discuss our results on ${La_{0.7}Sr_{0.3}MnO_{3}}$. Calculations based on the shell model predict four branches in the investigated energy range, all of oxygen character. Two of them are the bond-stretching and bond-bending branches that exist in the cubic perovskite structure. The other two are folded in by the tilt of the ${MnO_{6}}$ octahedra responsible for the rhombohedral structure. At the zone center one of the folded-in modes has a vanishing structure factor, whereas the other has a vanishing structure factor at the zone boundary. In agreement with the calculations we observe three phonons at the zone center at 42meV(bond bending), 53meV(folded in), and 72 meV(bond stretching). The bond-bending mode disperses sharply upwards, whereas the folded in mode is approximately flat as predicted by the calculation, and they cross near the reduced wavevector \textbf{q}=(0.15 0 0). The calculations predict an upwards dispersion of the bond-stretching branch in the (1 0 0) direction. Instead, we observe strong downward dispersion in agreement with previously published work \cite{Reichardt}. This anomaly has been observed in almost all metallic perovskite oxides and is generally interpreted as a signature of strong electron-phonon coupling. 
\begin{figure}[ptb]
%\centerline{\includegraphics[width= cm]{fig2.eps}} \caption{
\includegraphics[width=7 cm]{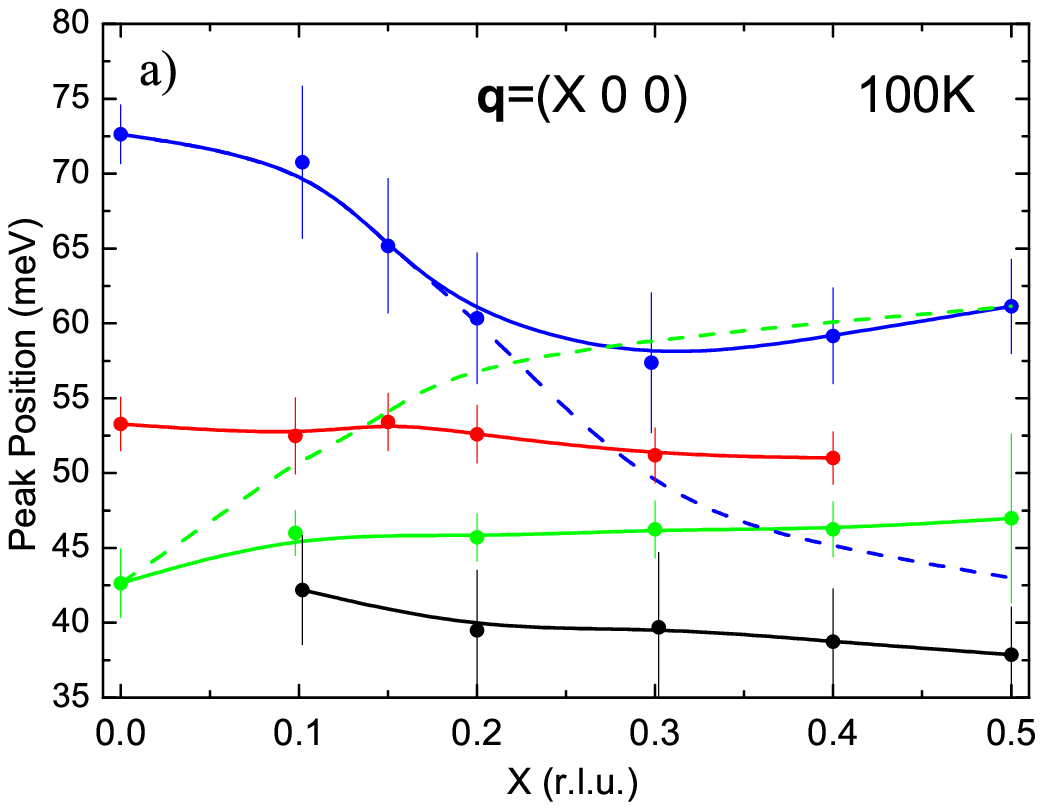}
\includegraphics[width=2 cm]{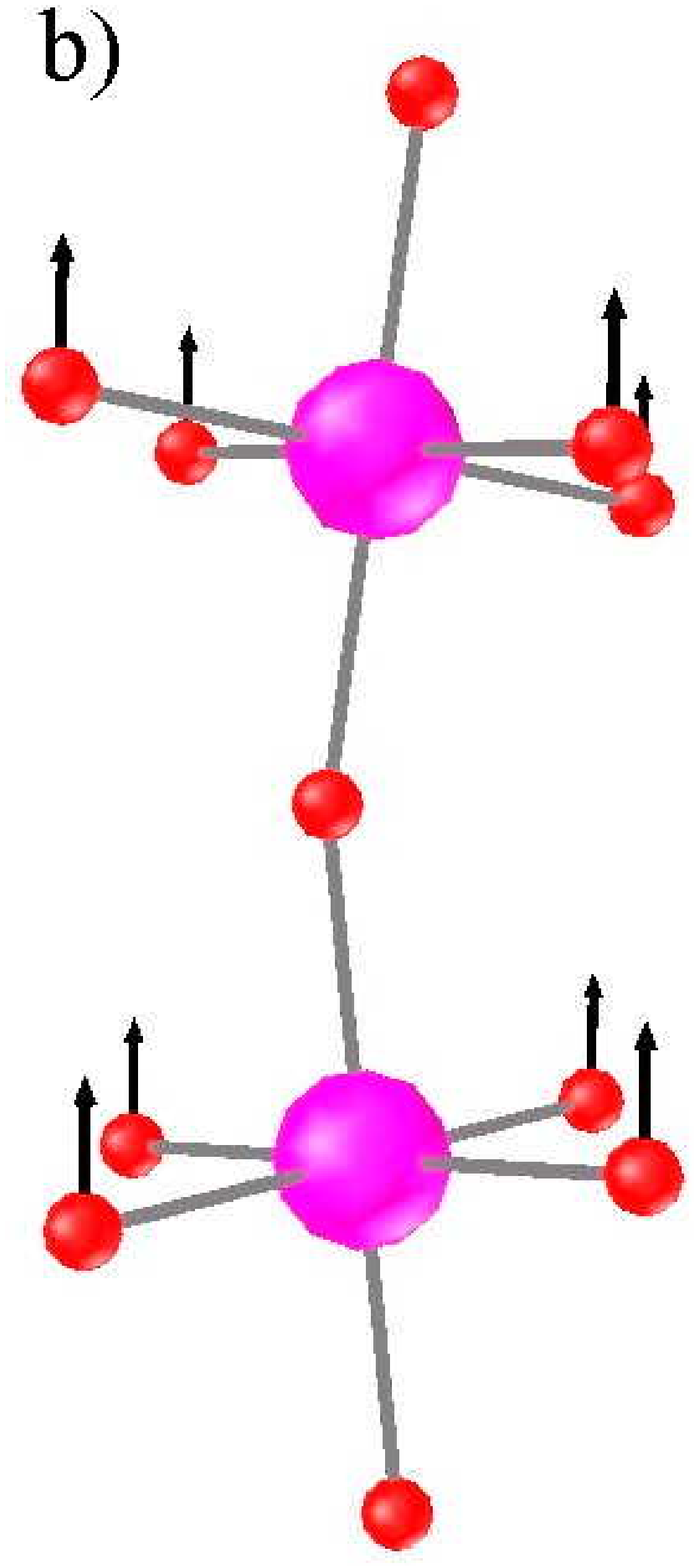} \includegraphics[width=2 cm]{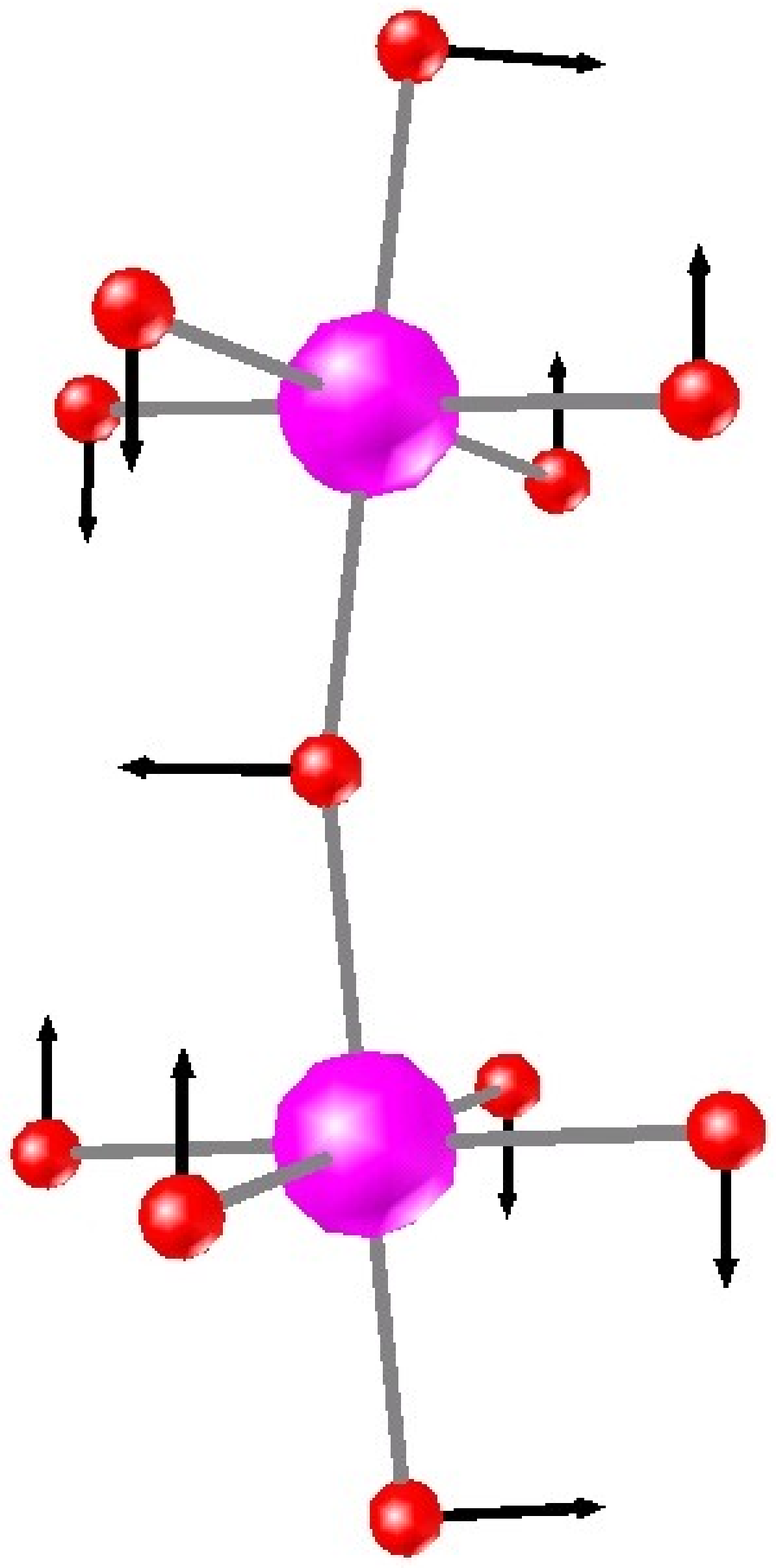} \includegraphics[width=2 cm]{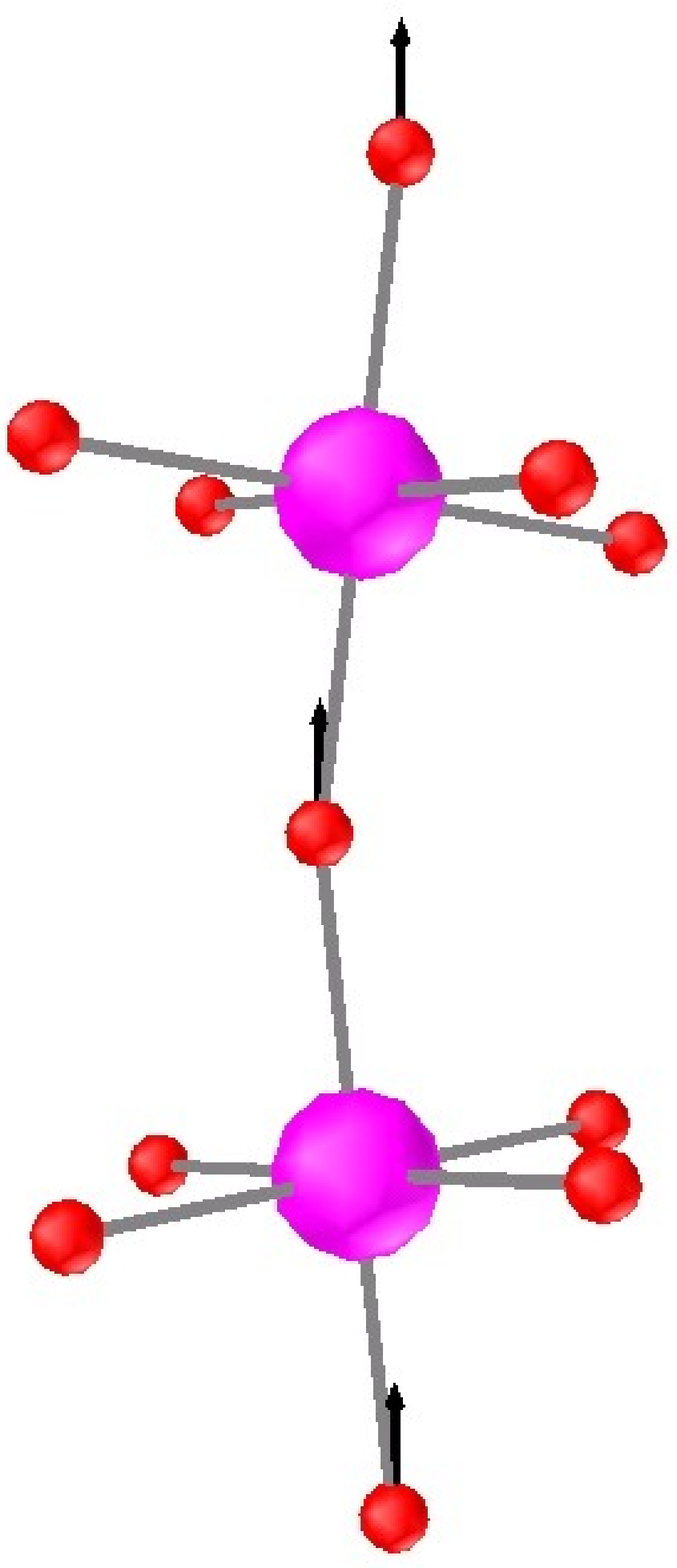}
\includegraphics[width=2 cm]{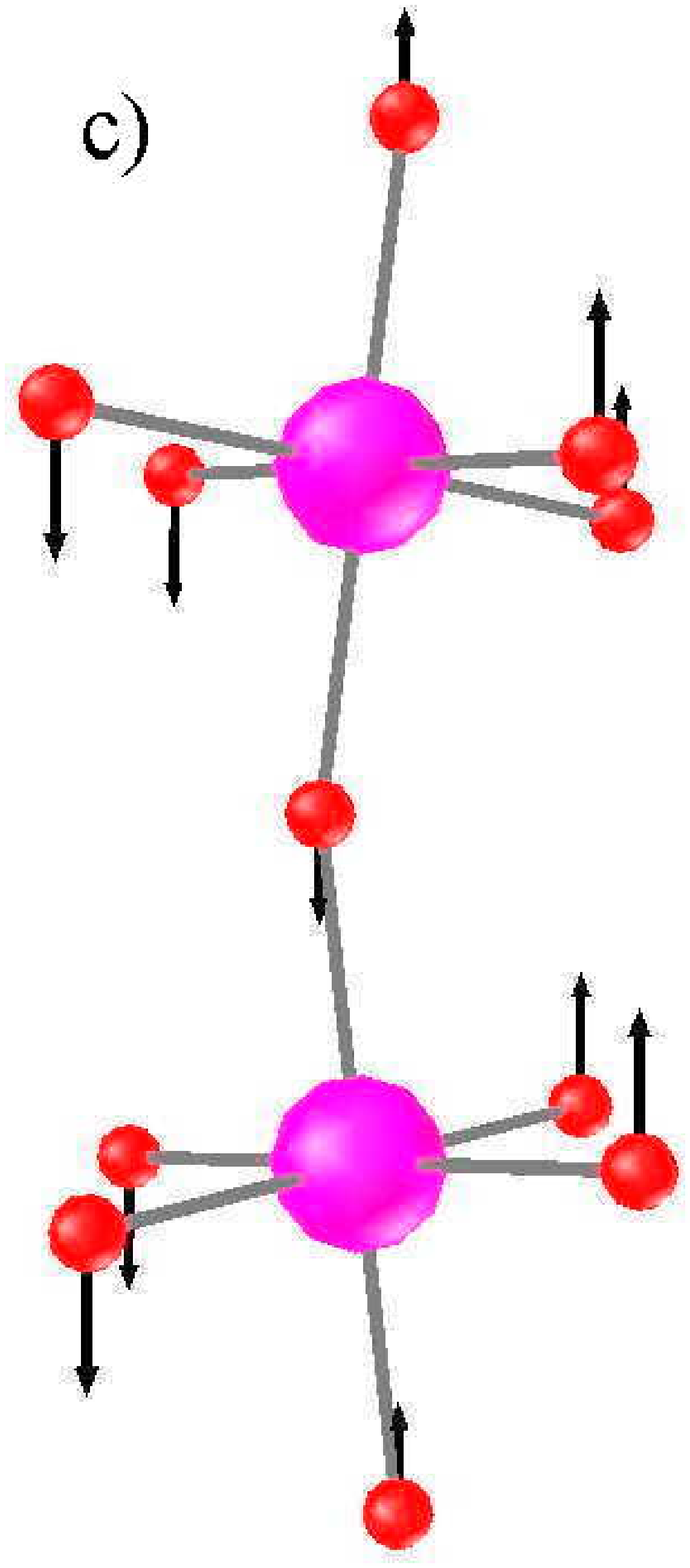} \includegraphics[width=2 cm]{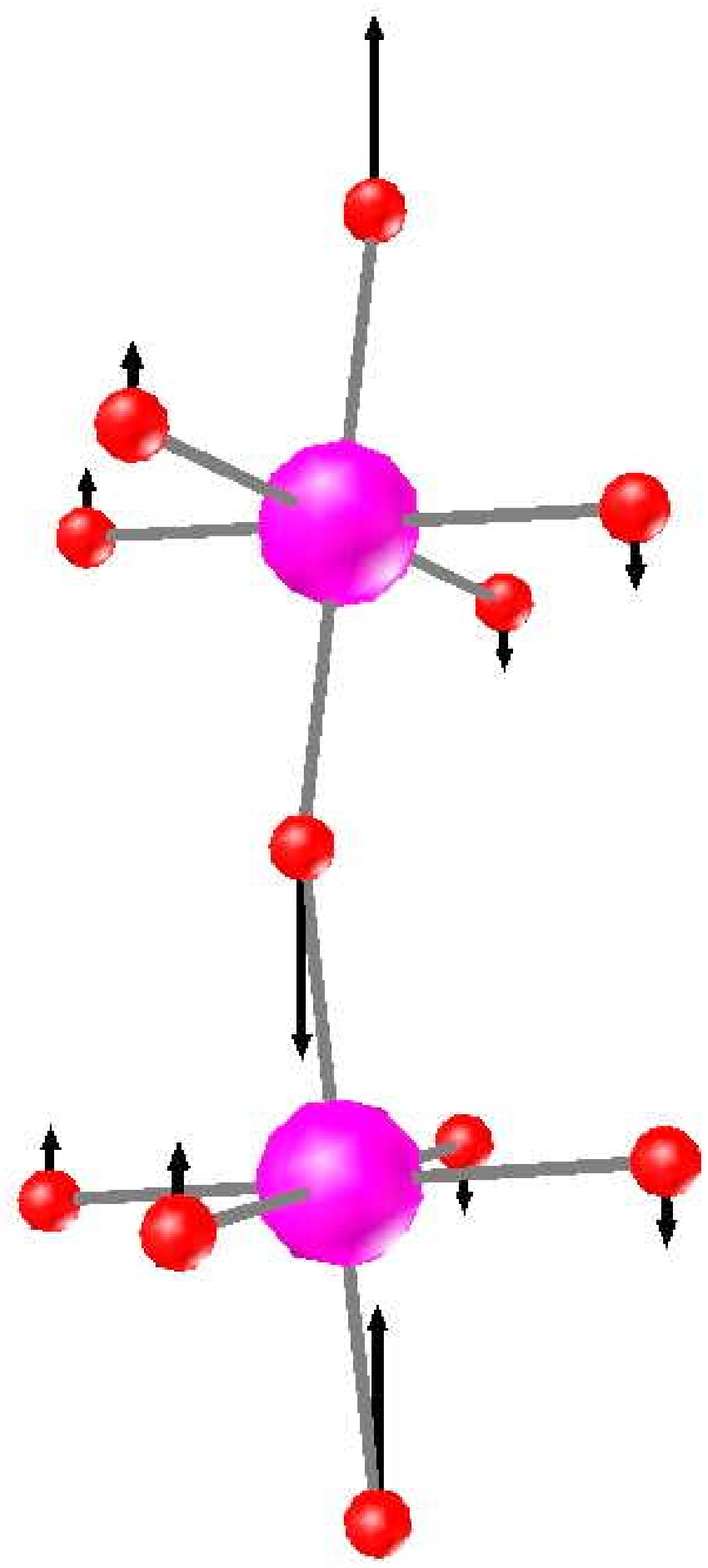} \includegraphics[width=2 cm]{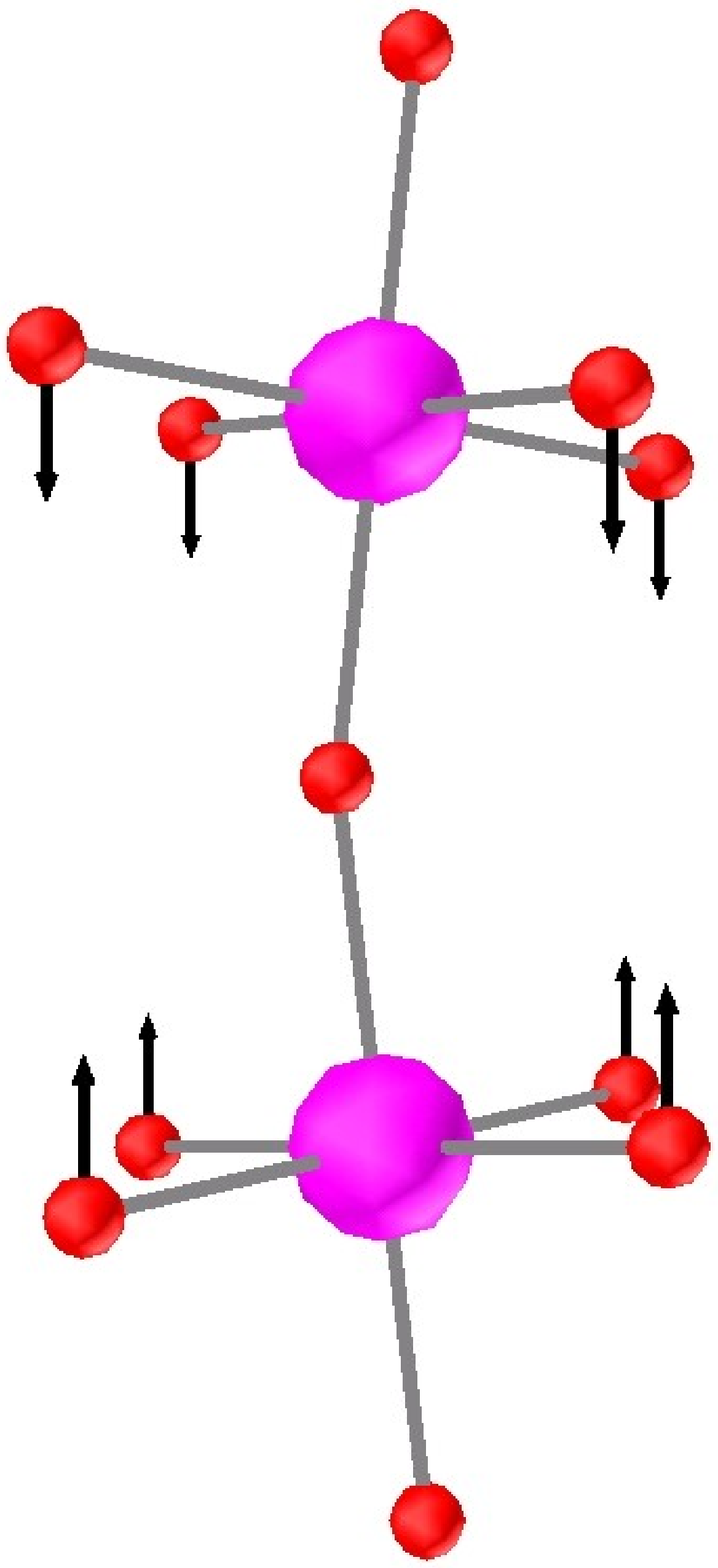} \caption{
{\label{fig1}} Phonon dispersions and selected approximate eigenvectors in ${La_{0.7}Sr_{0.3}MnO_{3}}$ a) Solid lines represent measured longitudinal phonon dispersions at 100K. Vertical bars represent peak widths. Mode polarizations change dramatically within each curve as a function of X. Dashed lines are guides to the eye showing dispersions of the bond-stretching and bond-bending character in the absence of interactions with other branches and each other. Their crossing of each other and of the folded in branches (see text) results in complex phonon polarization patterns.  b/c) Eigenvectors of the zone center/zone boundary modes at 100K projected onto the cubic (1 0 0) direction. Mode frequencies left to right in b/c are: 42/38mev, 53/47meV, 72/63meV.}
\end{figure}

The bond-stretching branch hybridizes with the folded-in branches at \textbf{q}=(0.2-0.5 0 0). (Fig. 1a) The hybridized modes have complex eigenvectors of mixed bond-bending and bond-stretching character and can be clearly identified by alternating strong/weak intensities in the Brillouin zones along the 100-direction. For example, the 47meV zone boundary phonon at 100K is observed at wavevectors \textbf{Q}=(2.5 0 0) and \textbf{Q}=(4.5 0 0) but not at \textbf{Q}=(3.5 0 0) and \textbf{Q}=(5.5 0 0). (Fig 2b,c) 
    
Taking advantage of the nearly cubic symmetry and using Brillouin zone dependence of the measured phonon intensities, we extracted approximate phonon eigenvectors at the zone center and the zone boundary (Fig. 1b). In the approximation we ignore deviations of the atomic displacements from the high symmetry cubic directions introduced by the rhombohedral distortion of the lattice. These should not be large, since the structure factors calculated from the eigenvectors in figure 1b agree well with experiment. At the zone center we have the bond-stretching, bond-bending, and a folded in bond-bending vibration at 72meV, 42meV, and 53meV respectively. At the zone boundary the "pure" bond-bending vibration is at 61 meV, and the modes at 39 meV and 47 meV are combinations of the bond-stretching mode and an out-of-phase bond-bending mode. (The bond-stretching mode and an out-of-phase bond-bending mode are coupled because of the rhombohedral lattice distortion.) Consistency between shell model results and approximate eigenvectors extracted from phonon intensities in different Brillouin zones further validates our analysis. 

\begin{figure}[t]
\centerline{\includegraphics[width= 9 cm]{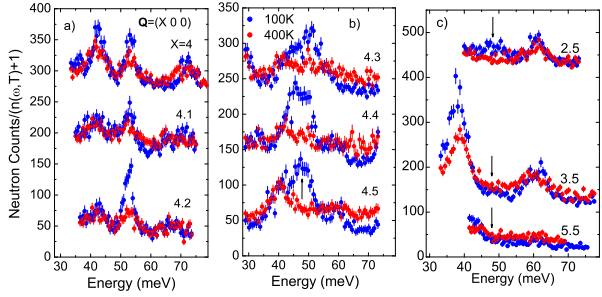}}\caption{
%\centerline{\includegraphics[width=\textwidth]{fig1.pdf}} \caption{}
{\label{fig2}} Phonon spectra in ${La_{0.7}Sr_{0.3}MnO_{3}}$ at 100K and 400K: a,b) Q=(4-4.5 0 0) c) Zone boundary spectra in other Brillouin zones. Arrows show the position of the 47 meV mode.}
\end{figure}

\begin{figure}[t]
\centerline{\includegraphics[width= 8 cm]{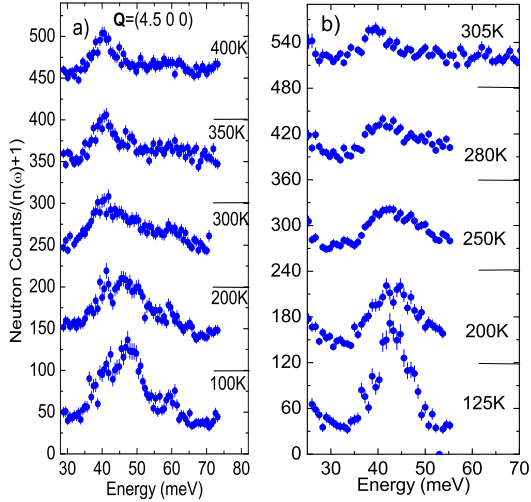}}\caption{
%\centerline{\includegraphics[width=\textwidth]{fig1.pdf}} \caption{}
{\label{fig3}} Temperature dependence of oxygen phonon response given by the neutron intensity divided by the Bose factor at Q=(4.5 0 0) in ${La_{0.7}Sr_{0.3}MnO_{3}}$ (a) and in ${La_{0.8}Sr_{0.2}MnO_{3}}$ (b)}
\end{figure}

Figures 2 and 3a illustrate strong temperature dependence of the phonon spectra in ${La_{0.7}Sr_{0.3}MnO_{3}}$. At the zone center the phonon intensities are consistent with the Debye-Waller factor. However, at the reduced wavevectors with q$_x>$0.15, a significant part of the combined spectral weight of the hybridized modes at 39 and 47 meV disappears on approach to the FM transition. Fits to the data show that the 39 meV peak decreases in intensity at \textbf{Q}=(3.5 0 0) and increases at \textbf{Q}=(4.5 0 0)(because its width increases) while the 47 meV peak disappears almost entirely above the FM transition. 

To test universality of the phonon anomaly, we also performed detailed measurements on ${La_{0.8}Sr_{0.2}MnO_{3}}$ at the zone center and at the zone boundary. The smaller O-Mn-O bond angle in this compound affects the structure factors of phonons with bond-bending character and shifts phonon frequencies, but the qualitative behavior does not change. At low temperature the zone center bond-stretching mode is at 70 meV and the zone boundary hybridized modes are at 44 and 35 meV. At low temperatures the 44 meV mode is very strong at \textbf{Q}=(4.5 0 0)(Fig. 3b) and an order of magnitude weaker at \textbf{Q}=(3.5 0 0), while the 35 meV mode is only observed at \textbf{Q}=(3.5 0 0)(data not shown). The temperature evolution is qualitatively similar to the one in ${La_{0.7}Sr_{0.3}MnO_{3}}$ scaled to the reduced transition temperature of 300K with an even greater loss of intensity from the higher energy mixed mode (Fig. 3). The modes move closer in energy as the temperature increases appearing at 37.5 and 39 meV at 310K. This effect may also exist in ${La_{0.7}Sr_{0.3}MnO_{3}}$ though it is not as clear from the data. 

Since the biggest intensity change is observed in the mode of predominantly bond-stretching character, the observed behavior is consistent with a significant loss of the bond-stretching component from the mixed modes above the transition. Calculations of eigenvectors based on the observed phonon intensiies above T$_c$ performed for ${La_{0.7}Sr_{0.3}MnO_{3}}$ confirm that very little bond-stretching character is left in the zone boundary phonons around 40 meV. At the same time the intensity of the zone center mode is consistent with the Debye-Waller factor. (Figs. 2a, 4b). Thus the loss of spectral weight away from the zone center cannot be explained by the loss of coherence, since in this case the entire branch should be suppressed independent of \textbf{Q}. 

\begin{figure}[t]
\centerline{\includegraphics[width= 8 cm]{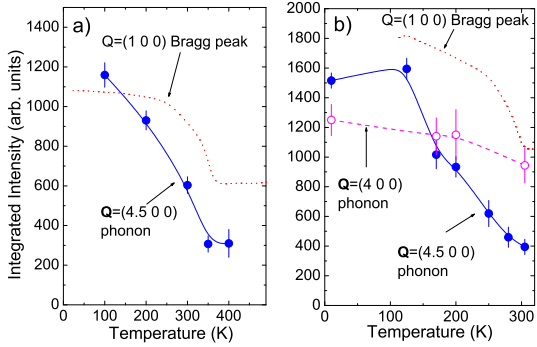}}\caption{
%\centerline{\includegraphics[width=\textwidth]{fig1.pdf}} \caption{}
{\label{fig4}} Temperature dependence of the integrated intensity of phonon peaks divided by the Bose factor compared with the intensity of the (1 0 0) Bragg peak (dotted line). Solid and dashed lines are guides to the eye. a) ${La_{0.7}Sr_{0.3}MnO_{3}}$, 47meV phonon at Q=(4.5 0 0) b) ${La_{0.8}Sr_{0.2}MnO_{3}}$, 44meV phonon at Q=(4.5 0 0) and 70 meV phonon at Q=(4 0 0)}
\end{figure}

\begin{figure}[t]
\centerline{\includegraphics[width= 8.5 cm]{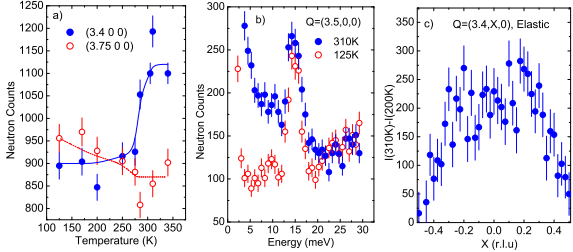}}\caption{
%\centerline{\includegraphics[width=\textwidth]{fig1.pdf}} \caption{}
{\label{fig5}} Results for ${La_{0.8}Sr_{0.2}MnO_{3}}$: a) Temperature dependence of elastic scattering at Q=(3.4 0 0) and Q=(3.75 0 0). b) Quasielastic scattering appearing at 310K near the zone boundary. c) Difference between elastic neutron intensities at 310K and 200K showing a broad polaron peak centered at X=0. Lines are guides to the eye.}
\end{figure}

Phonon intensity loss at \textbf{q}=(0.3 0 0) and (0.5 0 0) has also been seen in ${La_{0.7}Ca_{0.3}MnO_{3}}$ where it was interpreted as resulting from the loss of coherence of bond-stretching vibrations on approach to T$_c$ \cite{Zhang}. As discussed in the previous paragraph, this interpretation relies heavily on observing the intensity loss in the entire branch including the zone center. But in the data published in Ref \cite{Zhang} the zone center mode appears to be obscured by a temperature dependent contamination on the higher energy side. Until this ambiguity is removed, we conjecture that these results indicate a similar shift of spectral weight in ${La_{0.7}Ca_{0.3}MnO_{3}}$, thus the effect we observe is generic to ferromagnetic manganites.

Integrated spectral weight of the neutron scattering intensity obeys a sum rule, so the lost coherent intensity must shift to different energies at the same reduced reciprocal lattice vectors. There is no evidence for an intensity gain at higher energies except for the usual small increase in the background, thus the intensity must shift to lower energies. To look for the lost intensity there we performed several measurements on ${La_{0.8}Sr_{0.2}MnO_{3}}$ and found enhanced elastic intensity above the FM transition in the region of \textbf{q}-space where strong phonon renormalization occurs. Figure 5 illustrates this effect. At \textbf{Q}=(3.4 0 0) the elastic intensity increases sharply across the FM transition while there is no intensity enhancement at \textbf{Q}=(3.75 0 0), which is near the boundary of the region with the anomalous phonon effects. (A wavevector away from the zone boundary was chosen because \textbf{Q}=(3.5 0 0) is contaminated by multiple Bragg scattering.) In the transverse direction the intensity enhancement extends half way to the zone boundary (Fig. 5c). 

Following previous work, we assign the observed temperature-enhanced diffused scattering to correlated polarons. This polaron scattering appears in addition to the well known diffused peaks at \textbf{Q}=(3.75 ±0.25 0) that we observe as well. Preliminary measurements show that these polarons are dynamic as evidenced by observed quasielastic scattering (Fig. 5b), but the connection between the polarons and the quasielastic scattering needs to be confirmed.

According to the current theoretical picture, the electron-phonon coupling in the manganites occurs via the Jahn-Teller interaction. However, previous studies of Jahn-Teller phonons (which appear at \textbf{q}=(0.5, 0.5, 0) and (0.5, 0.5, 0.5)) show no strong effects on these modes \cite{Zhang}. So it is surprising to see such strong renormalization of the bond-stretching vibrations around \textbf{q}=(0.5, 0, 0), which might result from the frustration of the Jahn-Teller coupling. Arguriou et al. \cite{Argyriou} observed a soft mode in the bilayer manganite ${La_{2-2x}Ca_{2x}Mn_{2}O_{7}}$ at \textbf{q}=(0.25, 0.25, 0) around 23 meV, but it remains to be seen if there is any connection between this effect and our observations. 

Our results clearly indicate that despite the relatively small effect of T$_c$ on the electrical resistivity in the Sr-doped manganites, phonons play a key role in the FM transition. We also found evidence for polarons in the paramagnetic phase of these materials. The difference in the magnitude of MR between the Ca-doped and Sr-doped manganites possibly originates from the differences between polaron mobilities above T$_c$. It is important to note that in all cases phonon renormalization starts far below the onset temperature for the steep drop in the magnetic Bragg peak intensity indicating that the FM transition is always driven by the phonon anomaly. This result is entirely unexpected, since the currently accepted mechanism of the CMR assumes that it is the temperature-induced breakup of ferromagnetism that brings about the lattice effects.

The authors would like to thank L. Pintschovius for helpful discussions and critical readings of the manuscript. D. Reznik would like to thank M. Hennion, J.W. Lynn, D.N. Argyriou, and P. Dai for helpful discussions and suggestions.

\normalsize{$^\ast$To whom correspondence should be addressed;
E-mail: reznik@llb.saclay.cea.fr }

\end{document}